\documentclass[sigconf]{acmart}

\usepackage{makecell}
\usepackage{braket}
\usepackage{graphicx}
\usepackage{mathtools}
\usepackage{subcaption}
\usepackage{algorithm}
\usepackage{algorithmic}

\usepackage{fancyhdr}

\AtBeginDocument{%
  }

\copyrightyear{2023}
\acmYear{2023}
\setcopyright{rightsretained}
\acmConference[MICRO '23]{56th Annual IEEE/ACM International Symposium on Microarchitecture}{October 28-November 1, 2023}{Toronto, ON, Canada}
\acmBooktitle{56th Annual IEEE/ACM International Symposium on Microarchitecture (MICRO '23), October 28-November 1, 2023, Toronto, ON, Canada}
\acmDOI{10.1145/3613424.3614270}
\acmISBN{979-8-4007-0329-4/23/10}

\begin{document}

\title{Systems Architecture for Quantum Random Access Memory}

\author{Shifan Xu}
\affiliation{%
  \institution{Yale University}
  \department{Yale Quantum Institute}
  \city{New Haven}
  \state{CT}
  \postcode{06520-8263}
  \country{USA}}
\email{shifan.xu@yale.edu}

\author{Connor T. Hann}
\affiliation{%
  \institution{AWS Center for Quantum Computing\\ and\\ California Institute of Technology}
  \city{Pasadena}
  \state{CA}
  \postcode{91125}
  \country{USA}}

\author{Ben Foxman}
\affiliation{%
  \institution{Yale University}
  \department{Yale Quantum Institute}
  \city{New Haven}
  \state{CT}
  \postcode{06520-8263}
  \country{USA}}

\author{Steven M. Girvin}
\affiliation{%
  \institution{Yale University}
  \department{Yale Quantum Institute}
  \city{New Haven}
  \state{CT}
  \postcode{06520-8263}
  \country{USA}}

\author{Yongshan Ding}
\affiliation{%
  \institution{Yale University}
  \department{Yale Quantum Institute}
  \city{New Haven}
  \state{CT}
  \postcode{06520-8263}
  \country{USA}}
\email{yongshan.ding@yale.edu}


\begin{abstract}
  Operating on the principles of quantum mechanics, quantum algorithms hold the promise for solving problems that are beyond the reach of the best-available classical algorithms. An integral part of realizing such speedup is the implementation of quantum queries, which read data into forms that quantum computers can process. Quantum random access memory (QRAM) is a promising architecture for realizing quantum queries. However, implementing QRAM in practice poses significant challenges, including query latency, memory capacity and fault-tolerance. 

In this paper, we propose the first end-to-end system architecture for QRAM. First, we introduce a novel QRAM that hybridizes two existing implementations and achieves asymptotically superior scaling in space (qubit number) and time (circuit depth). Like in classical virtual memory, our construction enables queries to a virtual address space larger than what is actually available in hardware. Second, we present a compilation framework to synthesize, map, and schedule QRAM circuits on realistic hardware. For the first time, we demonstrate how to embed large-scale QRAM on a 2D Euclidean space, such as a 2D square grid layout, with minimal routing overhead. Third, we show how to leverage the intrinsic biased-noise resilience of the proposed QRAM for implementation on either Noisy Intermediate-Scale Quantum (NISQ) or Fault-Tolerant Quantum Computing (FTQC) hardware. Finally, we validate these results numerically via both classical simulation and quantum hardware experimentation. Our novel Feynman-path-based simulator allows for efficient simulation of noisy QRAM circuits at a larger scale than previously possible. Collectively, our results outline the set of software and hardware controls needed to implement practical QRAM.  
\end{abstract}



\maketitle

\section{Introduction}

Quantum computers hold the potential to solve problems that are beyond the reach of conventional digital computers. Such quantum speedup, as understood theoretically, arises from the utilization of quantum mechanical properties such as superposition and entanglement to process information more efficiently and rapidly \cite{nielsen2002quantum}. Some of the most promising quantum computing applications include quantum searching \cite{grover1996fast}, optimization problems \cite{van2020convex}, molecular simulation \cite{gilyen2019quantum, low2017optimal}, data processing for machine learning \cite{biamonte2017quantum, harrow2009quantum}, and cryptography \cite{shor1994algorithms}. For example, the quantum algorithm by Grover \cite{grover1996fast} for searching an unordered database of size $N$ makes only order of $\sqrt{N}$ queries to the database. This is a $\sqrt{N}$-speedup over the best classical algorithms, which require order of $N$ queries when given access to the same database.

Over the past three decades, technology for building quantum computing hardware has advanced steadily – prototypes of universal quantum processing units (QPU) housing 100+ individually programmable qubits are becoming available for the first time, and there is great interest in practically realizing these quantum applications. The development of scalable quantum computers is still in its early stages. Current Noisy Intermediate-Scale Quantum (NISQ) hardware \cite{preskill2018quantum} is limited by its system size (number of qubits) and fidelity (coherent lifetime of qubits and error rates of quantum gates). Remarkable progress \cite{byrd2023quantum} has been made in improving the performance of QPUs, through better quantum control, error correction architectures, as well as compiling and noise mitigation software. 

\begin{figure}[t]
    \centering
    \includegraphics[width=0.9\linewidth]{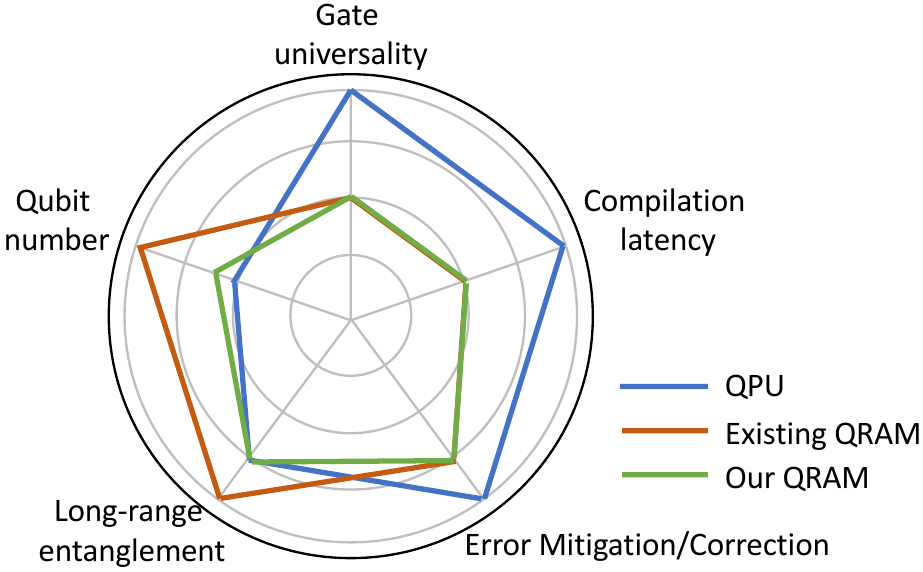} 
    \caption{Requirement map by quantum processor unit (QPU) and quantum random access memory (QRAM). There is a need for rethinking the systems architecture for QRAM because QRAM presents a distinct set of architectural constraints. A larger radial coordinate indicates a relatively more stringent requirement -- our QRAM design alleviates requirements in multiple dimensions. }
    \label{fig:motivation}
\end{figure}

One critical, yet largely missing, ingredient for realizing quantum speedup in practice is the implementation of \emph{quantum queries} \cite{nielsen2002quantum, biamonte2017quantum}, which allow data to be loaded into quantum states that the QPU can process. While QPU architectures are designed to process data rapidly, they often cannot encode classical data efficiently or robustly. This inhibits the practical deployment of many quantum algorithms. This issue is known as the \emph{data input and output (I/O) bottleneck} \cite{biamonte2017quantum, hann2021practicality}. In terms of gate and qubit overhead, the costs for encoding a large set of data into quantum states can be prohibitively high and dominate the costs of quantum algorithms. Traditional classical random access memory (RAM) \cite{jaeger1997microelectronic} allows data stored in memory cells to be loaded rapidly into a central processor unit (CPU). Similarly, a QRAM \cite{giovannetti2008quantum} has been proposed to enable quantum-mechanical loading of data into memory cells. This involves querying the QRAM simultaneously in a large superposition of different addresses.

More precisely, like in classical RAM, a memory cell can be accessed by specifying its address. That is, the data value $x_i$ is stored at address $i \in N$, where $N$ is the memory size.

\begin{align}
\text{Input: } i &\xrightarrow{\text{Classical RAM}} \text{Output: } x_i\\
\sum_{i=0}^{N-1}\alpha_i\ket{i}_{\text{A}}\ket{0}_{\text{B}} &\xrightarrow[]{\text{Quantum RAM}} \sum_{i=0}^{N-1}\alpha_i\ket{i}_{\text{A}}\ket{x_i}_{\text{B}}\label{eq:qram}
\end{align}
where $\alpha_i$ is the amplitude of each address in the superposition, and $\ket{\cdot}_{\text{A}}$ ($\ket{\cdot}_{\text{B}}$) is the address (bus) qubit register storing the input (output). 

While the design principles of a QRAM are similar to those of a classical RAM, there are unique challenges in scaling up QRAM in practice \cite{giovannetti2008architectures, hann2021practicality}: 
\emph{(i) Query Latency.} Like a RAM device, a QRAM allows any data to be accessed in almost the same amount of time regardless of the specified address. This includes a superposition of all addresses at once. Naively, entangling data from one memory cell at a time incurs latency, scaling with $N$ in the worst case. This latency can translate to a slowdown in the application, impeding its practical deployment. 
\emph{(ii) Memory Capacity.} Quantum algorithms, in principle, offer better quantum speedup when given access to large memory. However, existing QRAM architectures require a rapidly growing number of gates or ancillary qubits when scaling up the memory capacity.
\emph{(iii) Fault Tolerance.} Errors in the QRAM (e.g., due to various types of noise in the circuit) could seriously impact its utility, and in many cases eliminate the quantum advantage of an algorithm altogether \cite{regev2008impossibility}. As such, it is critical to guarantee the error robustness of QRAM, through either intrinsic noise resilience \cite{hann} or error correction.

We propose a general-purpose QRAM architecture to address these challenges, drawing insights from classical RAM, quantum compiling, and quantum error correction. Specifically, we make the following novel contributions:

\begin{enumerate}
    \item \textbf{Goal:} Small QRAM; large virtual memory.\\   \textbf{Solution:} We propose a new practical architecture that provides a virtual address space that can exceed the capacity of the physical QRAM. By hybridizing two previous query architectures \cite{babbush2018encoding, biamonte2017quantum}, 
    we achieve asymptotic savings in space and time.  As a result, our architecture enables queries to large memory that cannot be accomplished by either architecture alone. 
    \item \textbf{Goal:} Mapping qubits with minimal routing overhead.\\  \textbf{Solution:} QRAM requires strongly entangling $m$ address qubits with data from $O(2^m)$ memory cells. 
    
    Whether we can efficiently embed QRAM on practical (sparsely connected) two-dimensional hardware is highly non-trivial. For the first time, we provide a positive answer to this question. We present a constructive mapping of QRAM on 2D lattice architectures with minimal routing/communication overhead. 
    \item \textbf{Goal:} Noise-robustness of QRAM.\\ \textbf{Solution:} We show that our small-scale QRAM can be implemented on current hardware or near-term hardware with moderately improved error rates 
    We also demonstrate that small error correction codes allow us to substantially scale up QRAM with low overhead. \end{enumerate}

The rest of the article is organized as follows. In Sec.~\ref{sec:background}, we review the background on quantum compiling and architecture designs for QRAM. Sec.~\ref{sec:arch}  introduces a new QRAM architecture designed explicitly for hybrid QRAM. In Sec.~\ref{sec:mapping}, we present the algorithm for mapping QRAM on realistic hardware. In Sec.~\ref{sec:noise}, we analyze the biased-noise resilience property of our circuit and leverage it to reduce error correction overhead. Finally, we validate the results via classical simulation and quantum hardware experiments in Sec.~\ref{sec:eval} and Sec.~\ref{sec:results}. In Sec.~\ref{sec:results}, we also compare the resource usage of different QRAMs and show an asymptotic scaling advantage.

\section{Background}\label{sec:background}

\subsection{Principles of Quantum Computing}

In quantum computing, a quantum bit (qubit for short) is the fundamental computing unit. Unlike its classical counterpart, a qubit can be in a superposition state, that is a linear combination of $0$ and $1$. In the Dirac notation, $\ket{\psi}= \alpha\ket{0}+\beta\ket{1}$, where $\alpha,\beta$ are complex coefficients satisfying $|\alpha|^2+|\beta|^2 = 1$, and $\ket{0}=[1\;\; 0]^T$ and $\ket{1}=[0\;\;1]^T$ are the computational basis vectors. 
In quantum algorithms, quantum logic gates are used to manipulate the state of the qubits. Some common quantum logic gates are shown below:  

\begin{figure}[h]
    \centering
    \includegraphics[width=0.8\linewidth]{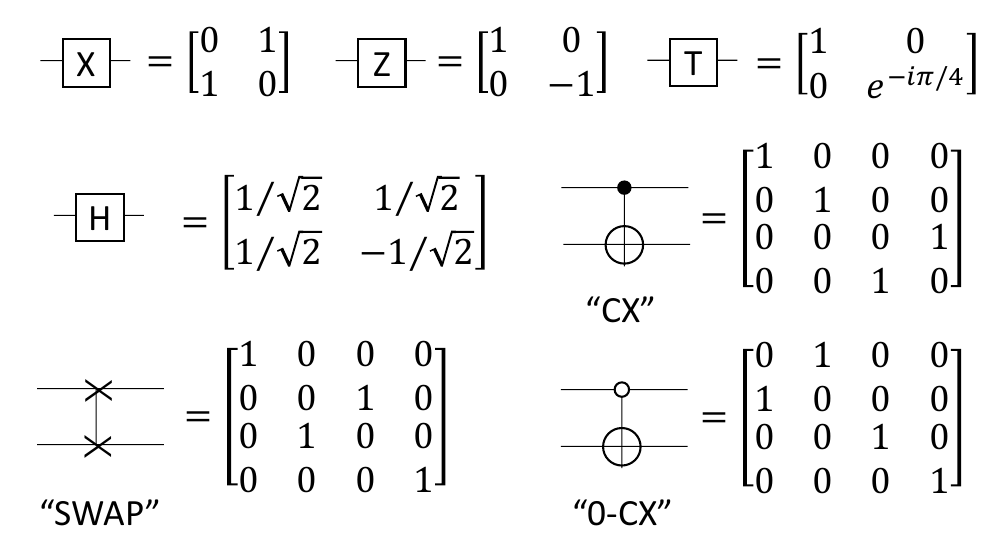}
\end{figure}

\subsection{Quantum Compiling}

A quantum compiler transforms a high-level quantum program or mathematical algorithm into a sequence of native instructions that the hardware backend recognizes \cite{chong2017programming, ding2020quantum}. The transformed quantum circuit must be logically equivalent, resource-efficient, and robust to hardware noise. Due to strict architectural constraints, compiler optimization will have to break traditional abstractions across the software stack and be adapted to algorithmic and device characteristics. We now highlight several important transformation passes in a compiler software.

\subsubsection{Gate Synthesis}

One of the first steps in quantum compiling is to decompose high-level unitary into a sequence of native gates from an instruction set \cite{harrow2001quantum}. A common instruction set is the Clifford+T gates, such as $\{\texttt{H}, \texttt{S}, \texttt{CX}, \texttt{T}\}$. When implementing quantum queries using these gates in practice, they can introduce latency into quantum algorithms if the circuit depth is too high.

As such, it is preferable to implement QRAM with a tailored gate set natively, such as the classical reversible gates, including \texttt{X}, \texttt{CX} (controlled-X), \texttt{Toffoli} (double-\\controlled-X), \texttt{MCX} (multi-controlled-X), and \texttt{CSWAP} (controlled-SWAP) gates. Otherwise, the multi-qubit gates can be decomposed into Clifford+T gates. For example, 
we can decompose a \texttt{CSWAP} gate to a circuit of depth 12, T depth 3, with no ancillae required \cite{Tdepth2,Tdepth3}. 

\subsubsection{Qubit Mapping and Routing}

To execute a quantum circuit, another critical step is to map all logical qubits onto the physical hardware. 
Current NISQ hardware has a limited number of qubits and restrictive qubit connectivity. Only adjacent qubits can interact with each other, while interactions between distant qubits are resolved via routing qubits closer to each other. Common routing strategies include physically moving qubits (e.g., for trapped ions) \cite{wu2021tilt} or logically swapping qubits (e.g., for superconducting circuits) \cite{murali2019noise}. Different routing strategies would incur different routing overhead, in terms of the number of additional operations. Future Fault-Tolerant Quantum Computing (FTQC) hardware will have similar constraints. For example, in surface codes, logical qubits can be laid out in a 2D grid topology. Logical gates and qubit routing can be accomplished via lattice surgery \cite{horsman2012surface}.

Although the general qubit mapping problem has been proven to be NP-hard \cite{mapnphard}, it is often useful to leverage information about the circuit and the hardware to improve the quality of a mapping strategy—a well-structured circuit can be easier to map and route \cite{zhang2021time, hillmich2021exploiting, molavi2022qubit}. Moreover, hardware noise-aware mapping strategies can enhance the circuit performance significantly  \cite{li2019tackling, tannu2019not, ding2020systematic,bhattacharjee2019muqut,murali2019noise}.

\subsection{Quantum Query Architectures}

\begin{figure*}[t]
    \centering
    \includegraphics[width=0.9\textwidth]{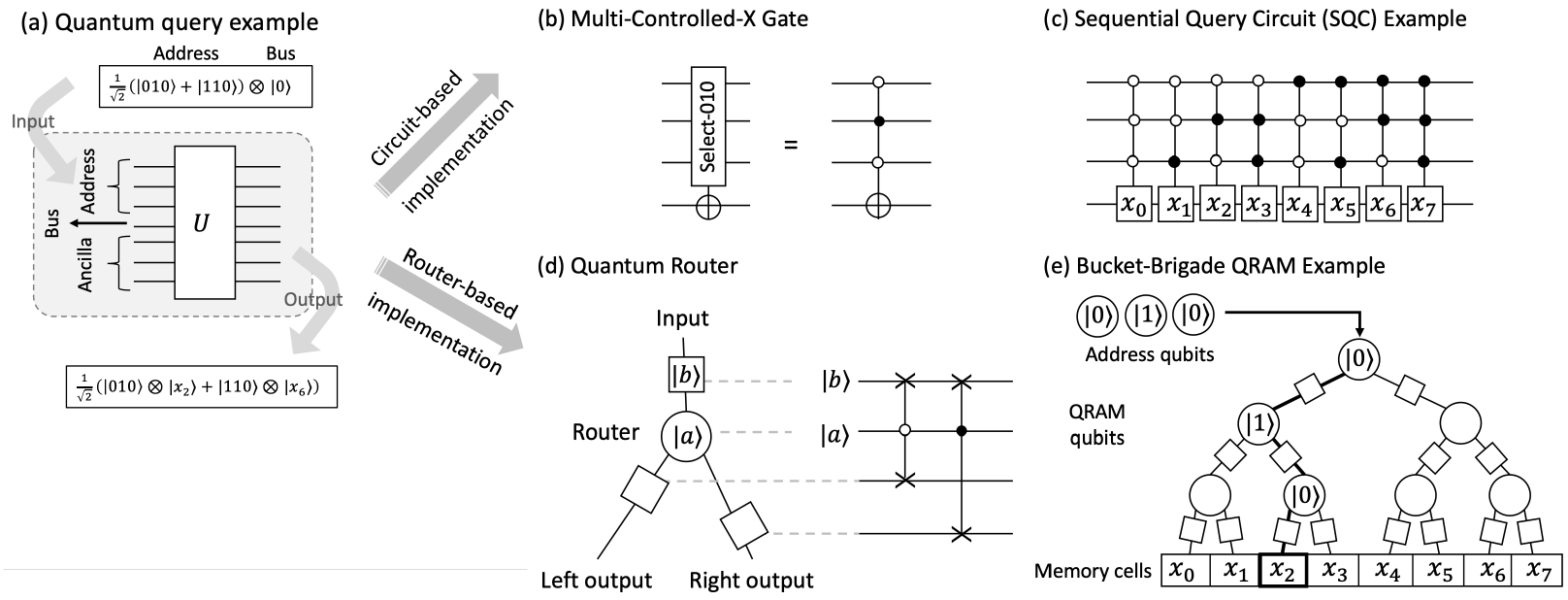}
    \caption{Different quantum query architectures and corresponding elementary units. (a) Quantum query example with address width 3. (b) Multiple-Controlled-X gate (MCX) as a unit for circuit-based query architecture. (c) Quantum router as a unit for router-based query architecture. (d) Sequential query circuit consisting of sequential MCX gates. (e) Parallel Bucket-Brigade QRAM consisting of recursive quantum routers.}
    \label{fig:sqc_router}
\end{figure*}

QRAM is an integral part of quantum computer architecture, as it enables quantum computers to efficiently encode classical data into a quantum state for the QPU to process. 
As shown in Equation~\ref{eq:qram}, a QRAM read operation therefore involves properly entangling data $x_i$ with the corresponding address $i$ in the superposition input.

\textbf{General-Purpose versus Domain-Specific} -- Query architectures can be categorized into two classes: (1) General-purpose (GP) architectures that load any possible data values $x_i$ for an arbitrary address $\sum_i\alpha_i\ket{i}$. (2) Domain-specific (DS) architectures that implement only particular data function(s) $f(i) = x_i$. DS architectures are useful when running applications in a target domain, as they are highly tailored to maximize efficiency or fault tolerance. 

We now describe the leading designs and implementations of QRAM, namely gate-based and router-based architectures. 
\subsubsection{Gate-Based Architecture}
Conventional wisdom is to use a partition of a universal QPU to implement the functionality of QRAM. As such, many proposals involve synthesizing the QRAM operation using a sequence of quantum logic gates and optionally using ancillary qubits.  
We provide two examples of these gate-based architectures below: Sequential Query Circuits and Reversible Logic Circuits.

\textbf{Sequential Query Circuit (SQC)} -- 
A quantum query can be implemented by a quantum circuit consisting of sequential \texttt{MCX} (multi-controlled-X) gates with no ancillary qubits. In the literature, the SQC is also known as a basic query circuit (BQC) or quantum read-only memory (QROM) \cite{babbush2018encoding}.  
Figure~\ref{fig:sqc_router}c provides a simple example of such a circuit. In this circuit, a sequence of $K$ \texttt{MCX} gates is applied to query a memory of size $N$, where each gate has $\log N$ controls on all the address qubits and one target on the so-called \emph{bus qubit} that holds the queried data.
Each \texttt{MCX} gate is responsible for loading data stored at one corresponding address, and the full query is realized by iterating sequentially over all possible addresses.
The SQC is a general-purpose architecture, capable of querying any function $f(i) =x_i$ for the memory cell data. It uses $O(\log N)$ qubits and has $O(N)$ query latency. 


\textbf{Reversible Logic Circuit (RLC)} -- 
An alternative implementation of quantum queries directly on a QPU using quantum gates is through an RLC. When the function that computes the data value is known, that is $x_i = f(i)$, one can synthesize this function directly with classical reversible gates (such as \texttt{X}, \texttt{CX} and \texttt{Toffoli}) and ancillary qubits. Because different circuits are required to implement different functions, RLC implements a domain-specific query. Classical reversible circuits are shown to be easier to optimize and verify than generic quantum circuits \cite{prasad2006data, ding2020square}. However, as with any DS architecture, the circuits must be synthesized and optimized for each domain application \cite{gidney2018halving}. This is useful if we want to implement a quantum computer to support a particular application. For example, the modular exponentiation step in Shor's algorithm \cite{shor1994algorithms}  
can be implemented by either an RLC or a hand-optimized quantum circuit \cite{gidney2021factor}. 

\subsubsection{Router-Based Architectures}
To minimize query latency, several router-based architectures are proposed. The defining feature of these general-purpose architectures is some form of \emph{quantum routing}, wherein quantum data is routed to multiple different locations in coherent superposition. The gadget in Figure~\ref{fig:sqc_router}d is a prototypical example; the gadget routes quantum data from an input to one of two different outputs conditioned on the state of a control quantum bit (router qubit). By leveraging such gadgets, router-based QRAMs can trade off space for time to implement $O(\log N)$-latency queries for a memory of capacity $N$, albeit at the cost of $O(N)$ qubits.  This exponential latency reduction relative to SQC is a significant step towards the practical implementation of QRAM. 
We provide two examples of these router-based architectures below: Fanout and Bucket-Brigade.

\textbf{Fanout QRAM}\cite{nielsen2002quantum} --
Fanout QRAM is the first architecture to achieve an $O(\log N)$-latency query. It arranges quantum routers in a binary tree, recursively using the outputs of the parent router as the inputs of the children routers. A query is implemented in two stages, address loading and data retrieval. During address loading, all routers are first initialized in $\ket{0}$, then a series of \texttt{CX} gates entangle the address and routers. In particular, all $2^k$ routers at level $k$ of the tree are flipped from $\ket{0}$ to $\ket{1}$ conditioned on the $k^\mathrm{th}$ address, resulting in the preparation of Greenberger–Horne–Zeilinger-like \\ (GHZ) states~\cite{greenberger1989going} across each level of the tree.
During data retrieval, a bus qubit is routed down from the root of the tree, following the path indicated by the routers' states. Once it reaches the bottom of the tree, classically-controlled gates copy the data $x_i = f(i)$ into the state of the bus. Subsequently, the bus qubit is back routed out of the tree, and the routers are returned to the all-$\ket{0}$ state by uncomputation. The fanout QRAM's $O(\log N)$ latency results from the fact that both the address loading and data retrieval can be implemented with only $O(\log N)$-depth circuits. However, Fanout QRAM is shown to suffer from decoherence problems due to the high entanglement of GHZ states \cite{hann}.

\textbf{Bucket-Brigade QRAM}\cite{giovannetti2008quantum, giovannetti2008architectures} --
Bucket-Brigade QRAM improves Fanout QRAM by modifying the address loading stage to reduce the entanglement among the routers, as shown in Figure~\ref{fig:sqc_router}e. Instead of using \texttt{CX} gates to entangle the address and routers, in Bucket-Brigade QRAM the address qubits are themselves routed into the tree, with the states of earlier address qubits controlling the routing of later ones. The resultant state of the routers is more akin to a \textit{W state} \cite{cabello2002bell} than a GHZ state.  The former has less entanglement entropy, which has been shown to greatly reduce the sensitivity of Bucket-Bridage QRAM to noise and errors \cite{hann}.  Importantly, $O(\log N)$ query latency is still achievable with this improvement, making it a competitive query architecture for the NISQ era.

\subsubsection{Other Architectures}

Other constructions of quantum queries have been proposed. For example, \textbf{Select-Swap QRAM} \cite{low2018trading} can be viewed as a combination of the gate-based and router-based architectures. Select-Swap QRAM employs a two-stage approach. During the first stage, one sequentially iterates over all possible states of a \emph{subset} of the address qubits, loading corresponding blocks of data for each. During the second stage, the remaining address qubits are used to route this data through a network of \texttt{CSWAP} gates, routing the queried data to a definite location. The sequential iteration in the first stage is analogous to the gate-based SQC, while the coherent routing in the second stage is reminiscent of the router-based architectures.

As another example, arbitrary state preparation algorithms can sometimes be used as a subroutine in QRAM. Some existing work includes the general unitary synthesis method \cite{zhang2004optimal,sun2023asymptotically} and parameterized circuit method \cite{phalak2022approximate,gokhale2019minimizing}. The general unitary synthesis is more complex than router-based QRAM, with specific gates for the different classical data sets. This increases the difficulty of changing the classical dataset. The parameterized circuit is a popular NISQ-era application that has been proposed to have the ability to realize approximate quantum query with $O(1)$ depth and $O(\log N)$ number of qubits with the price of long training time and approximate queries.

\section{Proposed QRAM architecture}
\label{sec:arch}

\begin{figure}[t]
    \centering
    \includegraphics[width=0.9\linewidth]{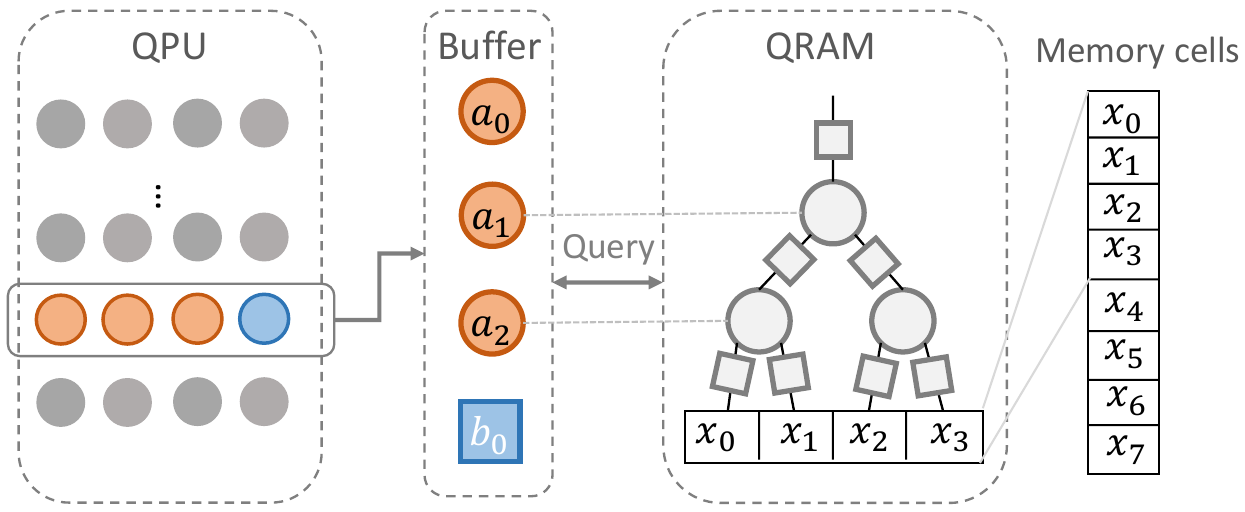}
    \caption{Overview of the proposed virtual QRAM architecture ($k=1, m=2$) and its interaction with QPU. The QPU qubits are swapped to the buffer for a quantum query and returned to QPU once the query is done by QRAM.}
    \label{fig:qpu-qram}
\end{figure}

This work demonstrates an end-to-end architecture design for QRAM. The prior works, including Bucket-Brigade QRAM and FANOUT, have unaffordable overhead in terms of resource consumption when querying a large database. Our proposed system architecture is built upon a novel router-based construct that allows us to query a virtual address space larger than that is physically available, as shown in Section~\ref{subsec:query}. We also provide a series of optimizations in Section~\ref{subsec:optimizations} to significantly cut down both space and time costs. 
Our new design decouples the address-loading stage from the data-retrieving stage in a quantum query, and deeply optimizes each stage. As a result, we can achieve an asymptotic advantage over prior work (which is shown in Section~\ref{sec:results}). 

\subsection{Two-Stage Query Overview}
\label{subsec:query}

We introduce a novel router-based QRAM using $O(M)$ qubits, where $M = 2^m$. We define the \emph{address width} $m$ as the number of bits to specify an address and the capacity $M$ the size of the address space. For illustration purposes, we include an example architecture shown in Figure~\ref{fig:qpu-qram}, and a step-by-step query procedure outlined in Figure~\ref{fig:query}, with the detailed circuits shown in Figure~\ref{fig:circuits}. Like in the router-based QRAM architectures, our construction consists of two stages: address loading and data retrieving. We explain these two stages in turn below.

\subsubsection{Stage 1: Address Loading}
The task in the address loading stage is to route the $m$ address qubits into the small QRAM. For this purpose, we follow the approach of other router-based QRAM architectures and arrange a collection of quantum routers in a binary tree structure. We affix a layer of $M$ \emph{data qubits} to the outputs of routers at the lowest level of the tree, with each data qubit corresponding to one of the $M$ classical memory cells. These data qubits will be used to facilitate operations during the data retrieval stage.

The address-loading stage follows the conventional Bucket-Brigade procedure: the address qubits are sequentially routed into the tree, with the states of earlier address qubits controlling the routing of later ones. The only difference is that, in our scheme, the data qubits are subsequently prepared in a special state conditioned on the states of the routers (Figure~\ref{fig:query}a). Specifically, when address $i$ is queried, the $i^\mathrm{th}$ data qubit is flipped from $|0\rangle$ to $|1\rangle$. This flipping is implemented via a collection of \texttt{CX} gates, with the bottom layer of routers as controls and the data qubits as targets, see Fig~\ref{fig:circuits}a as query state preparation.

\subsubsection{Stage 2: Data Retrieval}
We propose a \emph{novel} data-retrieval stage, which, at a high level, performs data compression from the bottom data qubits to the root data qubit in the QRAM. First, $M$ classically-controlled gates act on the bottom of the tree to write $\ket{x_i}$ on the data qubits when address $\ket{i}$ is queried. Specifically, each data qubit is paired with an ancillary qubit initialized in $\ket{0}$. We refer to the two-qubit system of a data qubit and its ancilla as a data node. Then, conditioned on $x_i$, a \texttt{SWAP} gate is applied between the two physical qubits in a data node. This has the effect of encoding the classical data in a dual-rail encoding, i.e.~$x_i = 0$ (resp.~$x_i = 1$) is encoded as $\ket{10}$ ($\ket{01}$). Fig.~\ref{fig:query}(b) shows the resultant state.

Then, an array of \texttt{CX} gates (Figu~\ref{fig:circuits}b) is used to propagate this data up to the root node of the tree, as shown schematically in Fig.~\ref{fig:query}c. Next, the data at the root node is copied to the bus qubit, conditioned on the $k$ remaining address qubits, using a \texttt{MCX} gate. After that, the \texttt{CX} array is applied again to uncompute and disentangle QRAM qubits, returning the QRAM to the state for the next data retrieval stage (Fig.~\ref{fig:query}d).\ We can then repeat the data retrieval stage for the next segment/page of the memory, after swapping the new segment into the bottom of the tree. Crucially, in this new data-retrieval step, the only non-Clifford gate involved is the \texttt{MCX} gate.

\subsubsection{Putting It Together: Virtual QRAM}
\label{subsec:virtual}

The goal is to implement quantum query access to a memory of capacity $N = 2^n$, where $N>M$. That is, we consider a practical scenario where $O(2^n)$ qubits are not physically available. Can we still implement the query, and if so, at what cost? This scenario resembles classical memory architecture design where ``virtual memory'' allows a small physical RAM can access a large address space by swapping segments (also known as pages) of memory from disk storage \cite{kilburn1962one}. Our ``virtual QRAM'' design follows precisely this intuition: we implement a virtual QRAM that allows a small QRAM with $O(M)$ qubits to access a large address space $O(N)$ by swapping segments of classical memory, as shown in the right panel of Fig.~\ref{fig:qpu-qram}. To accomplish this, we need to design a QRAM architecture that allows us to query segments of memory coherently.

\begin{figure}[t]
         \centering
         \includegraphics[width=0.8\linewidth]{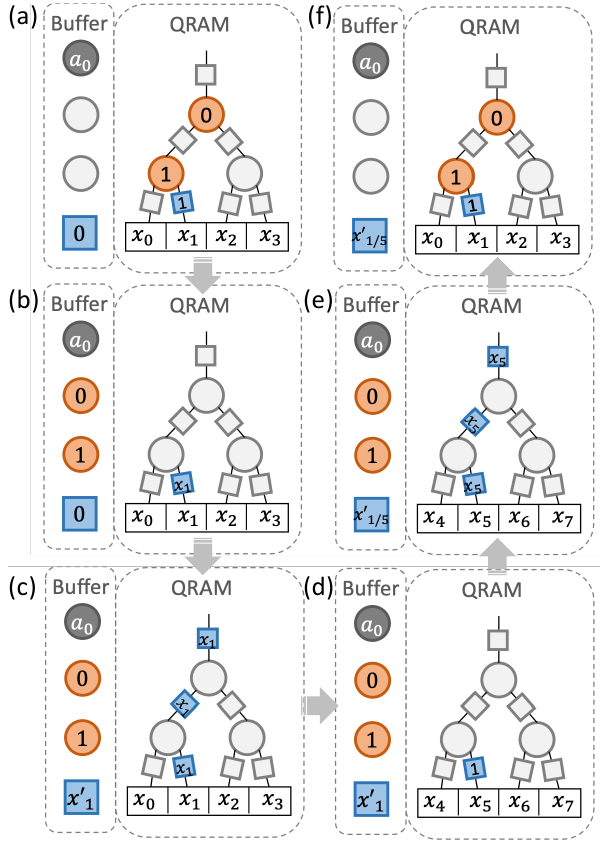}
         \caption{The step-by-step procedure of a quantum query to virtual QRAM ($k=1,m=2$). (a) Load the $m$ address qubits (orange) in QRAM and prepare the specific data qubit (blue) for data retrieval. (b) Coherently write classical data $x_i$ in data qubits to state $\ket{x_i}$. (c) Retrieve data by copying to root qubit via \texttt{CX} gates. Data is copied to the bus qubit conditioned on the state of $k$ address qubits (dark gray). (d) Uncompute data retrieval and swap in next segment memory. (e) Repeat data retrieval in (c). (f) Uncompute data retrieval.}
         \label{fig:query}
\end{figure}

\begin{figure*}[t]
     \centering
         \centering
         \includegraphics[width=\textwidth]{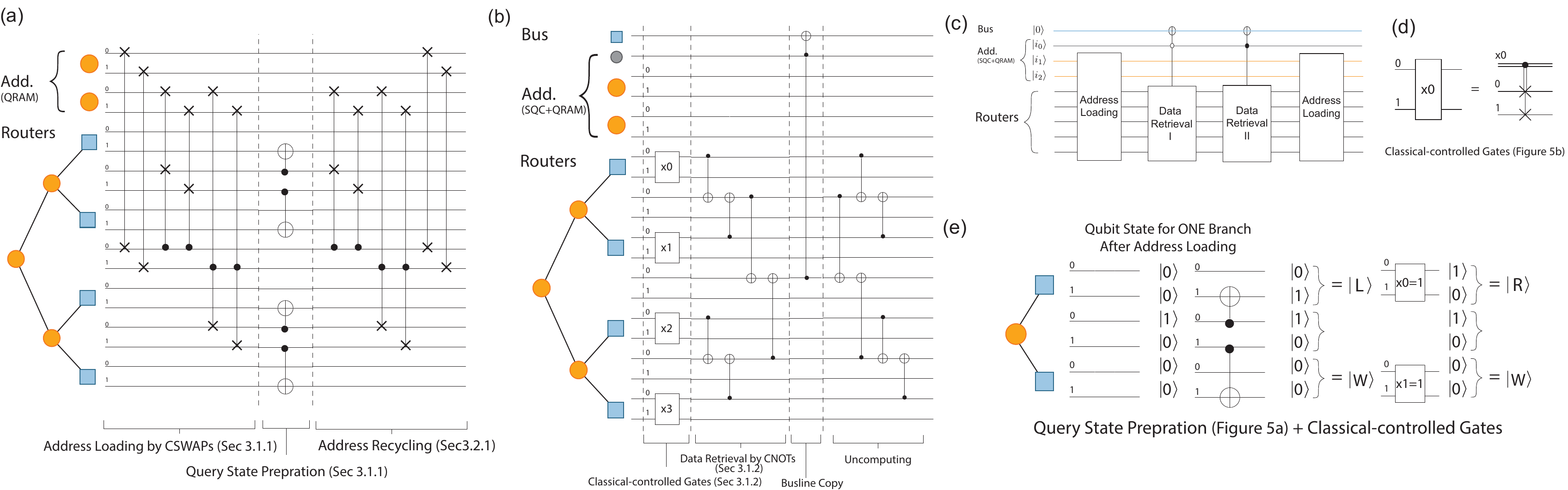}
         \caption{Circuit for the architecture design of virtual QRAM, with SQC width $k=1$ and QRAM width $m=2$. (a) Circuit for Address Loading Stage. (b) Circuit for Data Retrieval Stage.  (c) Outline of a full virtual QRAM circuit. (d) Classical-controlled SWAP gates with dual-rail encoding \ (e) Qubit state in query state preparation and data retrieval.}
         \label{fig:circuits}
\end{figure*}

We first partition the full size-$N$ classical memory into $K=2^k$ continuous segments (pages), each of which has $M=2^m$ memory cells. It is equivalent to viewing the original $n$-bit address as two parts: the most significant $k$-bits (which we call SQC width) and least-significant $m$-bits (which we call QRAM width), where $k+m=n$. Our design hybridizes gate-based SQC and router-based QRAM, where a quantum query consists of 6 basic steps: (a) loading $m$ address bits into QRAM, (b) preparing leaf data qubits for data retrieval, (c) retrieving data to root qubit, (d) uncomputing data retrieval, (e) repeat (b-d) for each segment of classical memory, (f) uncomputing address loading. More details are illustrated in Figure~\ref{fig:query}. A distinct feature of our design is the ``load-once'' property. Our method only loads the $m$ address qubits into QRAM once at the beginning (and at the end for uncomputation) as shown in Figure~\ref{fig:circuits}c, whereas in a previous design from \cite{hann} the $m$ address qubits need to be loaded $2^k$ times. This is one of the major sources of gate savings in our method. We present a pseudocode algorithm to describe the entire query procedure in Algorithm~\ref{alg:query}. Note that, for illustration purposes, this algorithm does not include the optimization techniques introduced in Section~\ref{subsec:optimizations}.

\begin{algorithm}[t]
\caption{Quantum Query with Virtual QRAM.}\label{alg:query}
\begin{algorithmic}
\REQUIRE $\ket{\psi_A} = \sum \alpha_i\ket{i}$,$ n \geq 1 , 0 \leq k \leq n-1 , m = n-k$
\ENSURE $\ket{\psi_{AB}} = \sum_{i=0}^{N-1}\alpha_i\ket{i}_A\ket{x_i}_B$
\STATE $q_t^{(a)}, q^{(b)}$ $\gets$ $t_{th}$ address qubit, Bus qubit 
\STATE $q_t^{(c)}, q_t^{(d)}$ $\gets$ $t_{th}$ layer router qubit, data qubit
\STATE $q_{-1}^{(d)}$ $\gets$ input qubit to the root router
\STATE SWAP[$q_0^{(a)} , q_{-1}^{(d)}$], SWAP[$q_{-1}^{(d)}, q_0^{(c)}$] \COMMENT{Address Loading}
\FOR{$u\gets1\dotsc m-1$} 
\STATE SWAP[$q_u^{(a)} , q_{-1}^{(d)}$]
\FOR{$v\gets0\dotsc u-1$} 
\STATE CSWAP[$q_v^{(c)} , q_{v-1}^{(d)} , q_v^{(d)}$]
\ENDFOR
\STATE SWAP[$q_{u-1}^{(d)}, q_u^{(c)}$]
\ENDFOR
\STATE CX[$q_{m-1}^{(c)} , q_{m-1}^{(d)}$]
\FOR{$ p \gets0\dotsc  2^k-1 $}  
\STATE Classical-CX[$x_i, q_{m-1}^{(d)}$] \COMMENT{Data Retrieval}
\FOR{$u\gets0,m-1$} 
\STATE CX[$q_{m-1-u}^{(d)} , q_{m-2-u}^{(d)}$]
\ENDFOR
\STATE MCX[$q_{m,\dots,n-1}$ , $q_{-1}^{(d)}$, $q^{(b)}$]
\FOR{$u\gets0\dotsc m-1$}   
\STATE CX[$q_{m-1-u}^{(d)} , q_{m-2-u}^{(d)}$] \COMMENT{Uncompute: Data Retrieval}
\STATE Classical-CX[$x_i, q_{m-1}^{(d)}$]
\ENDFOR
\ENDFOR
\STATE CX[$q_{m-1}^{(c)} , q_{m-1}^{(d)}$] \COMMENT{Uncompute: Address Loading}
\FOR{$u\gets1\dotsc m-1$} 
\STATE SWAP[$q_u^{(a)} , q_{-1}^{(d)}$]
\FOR{$v\gets0\dotsc u-1$} 
\STATE CSWAP[$q_v^{(c)} , q_{v-1}^{(d)} , q_v^{(d)}$]
\ENDFOR
\STATE SWAP[$q_{u-1}^{(d)}, q_u^{(c)}$]
\ENDFOR
\STATE SWAP[$q_{-1}^{(d)}, q_0^{(c)}$], SWAP[$q_0^{(a)} , q_{-1}^{(d)}$]
\end{algorithmic}
\end{algorithm}

\subsection{Implementation and Optimizations}
\label{subsec:optimizations}

\subsubsection{Key Optimization 1: Address Qubit Recycling}

As shown in Figure~\ref{fig:query}, the internal router qubits (orange nodes) are not being used during steps (b)-(e). By recycling/reusing these qubits in replacement of the data qubits, we do not need any data qubits (blue nodes) for the quantum routers internal to the tree. 
Figure \ref{fig:circuits} illustrates that the data retrieval stage reuses router qubits for copying data via the \texttt{CX} array. As such, in diagrams such as Figure~\ref{fig:teleportation}c and Figure~\ref{fig:biasedz}, we do not draw blue data qubits for internal quantum routers.

\subsubsection{Key Optimization 2: Lazy Data Swapping}
As shown in the data-retrieval stage in Figure \ref{fig:circuits}b, classical data is loaded and unloaded sequentially for each segment of classical memory. We observed that if the subsequent classical data corresponding to the memory address on the next page is equal to the previous one, unloading and reloading data qubits are redundant and unnecessary. Instead, by computing $x_i'=x_i\bigoplus x_{i+2^m}$, it is necessary to load the next classical data $x_i$ only when $x_i'=1$. At the final data-retrieving stage, an alternative classical data unloading is accepted, with a classical value as $c_i=\bigoplus_{i\in \{0,1,\dots,k\}} x_i$. Adopting this technique, named lazy data swapping, provides \texttt{SWAP} gate savings of $O(2^{n-1})$ in average cases, since the subsequent classical data can be the same as the original data with a probability of $p=0.5$, assuming a uniform distribution for the classical data $x_i$.

\subsubsection{Key Optimization 3: Address Pipelining}
With pipelining, we can reduce the depth of address-loading from $O(m^2)$ to $O(m)$. In the naive approach to address loading, address qubits are routed into the tree sequentially, with the $(\ell+1)^\mathrm{th}$ address qubit \emph{waiting} to be routed until the $\ell^\mathrm{th}$ address qubit has reached its destination at level $\ell$ of the tree. The total routing time is thus $\sim \sum_{\ell=1}^m \ell = O(m^2)$.  
Instead, the addresses can be routed in a pipelined manner: the $(\ell+1)^\mathrm{th}$ address qubit is routed into the tree immediately after the $\ell^\mathrm{th}$ qubit has been routed one layer down, i.e.~without waiting. Removing the waiting reduces the depth to $O(m)$. While the resulting circuit is equivalent to the parallel schedules introduced in \cite{hann} and \cite{di2020fault}, we identify the origin of such parallelism as coming from pipelining.

\section{Mapping QRAM in 2 Dimensions}
\label{sec:mapping}
\subsection{Mapping and Routing: Challenges}

\begin{figure*}[t]
         \centering
         \includegraphics[width=0.95\textwidth]{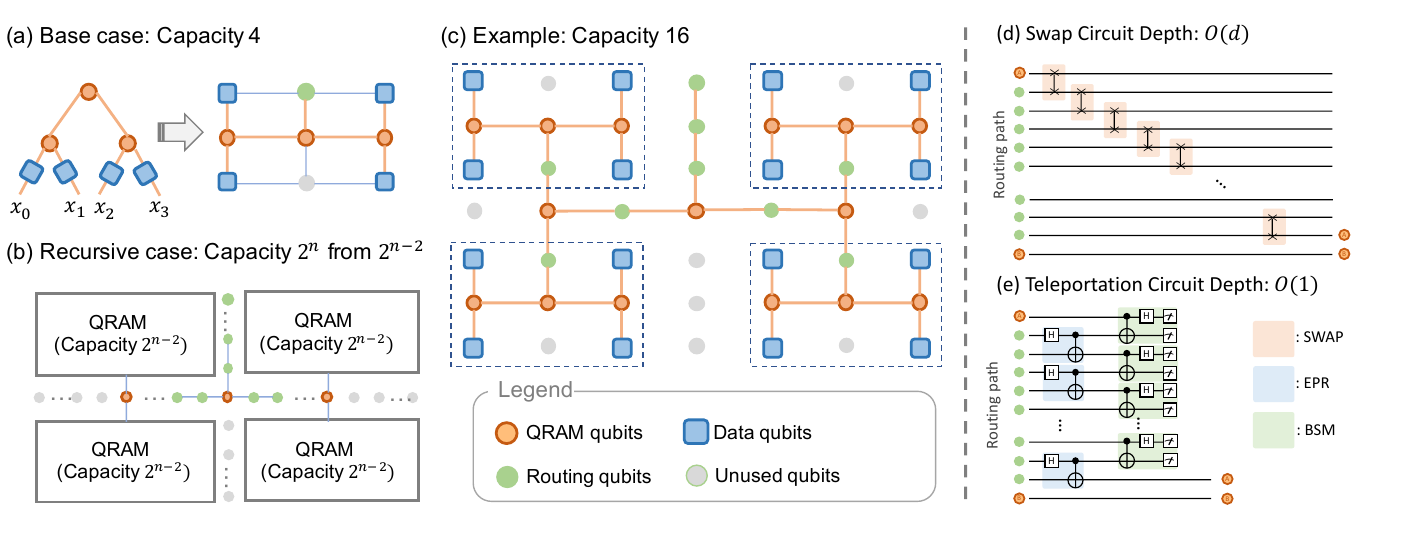}
         \caption{(a) H-Tree embedding of a capacity-4 QRAM on a 2D grid. (b) Recursive H-Tree embedding for a capacity-$n$ QRAM (Sec.~\ref{subsec:htree}). (c) Example embedding of a capacity-16 QRAM. (d) Routing via swapping. (e) Routing via teleportation (Sec.~\ref{subsec:teleport}). }
         \label{fig:teleportation}
\end{figure*}

Qubit mapping and routing are important steps in a quantum compiler to implement a quantum algorithm on hardware. Routing overhead refers to the number of operations needed to execute a gate operation on two (possibly physically distant) qubits. This overhead can cause a significant increase in the query latency of the quantum algorithm.

Mapping QRAM (of capacity $M$) is particularly challenging because it is required to map and entangle $O(M)$ qubits. The tree-like structure in router-based QRAM means that the $i^{\text{th}}$ layer involves $O(2^i)$ qubits and multi-qubit gates. Naively, keeping a router qubit in layer $i$ equal-distance with its parent router in layer $i-1$ and its children routers in layer $i+1$ is only possible in hyperbolic geometry. To embed a tree in 2D Euclidean space, the root ($i=0$) of the tree can be far apart from the next layer down ($i=1$) due to the large size of the two subtrees. 

Our research shows that QRAM can be embedded in a 2D nearest-neighbor grid without incurring asymptotic routing overhead. In other words, we can map the QRAM circuit on a 2D grid and route the qubits without increasing the $O(\log M)$ depth of the original circuit. This is achieved by combining a mapping strategy via topological minor graph embedding (Sec~\ref{subsec:htree}) and a routing method based on teleportation (Sec~\ref{subsec:teleport}).

\subsection{Mapping QRAM via H-Tree Recursion}
\label{subsec:htree}

To map QRAM onto a 2D architecture, we need to find an embedding of a binary tree in the connectivity graph of the hardware. In addition, we require the embedding to be a \emph{topological minor graph} embedding. This allows us to implement the teleportation-based routing method by ensuring all routing qubits do not carry any logical information. Given a simple, undirected graph $G$, another graph $H$ is a topological minor of $G$ if $H$ can be obtained from a subgraph of $G$ by deleting edges, vertices, and contracting edges. 

We reduce the problem of mapping QRAM to embedding a complete binary tree in a 2D grid. The problem of embedding complete binary trees into grids has been extensively studied in classical VLSI design \cite{heckmann1992optimal,opatrny2000embeddings,lin2003expansion,ducourthial1998implementing}. The H-tree recursion is the first efficient mapping strategy introduced by \cite{browning1980tree}. In Figure \ref{fig:teleportation}a, we present the optimal embedding of $T_2$ into Grid$(3,3)$  by H-tree recursion, which is also the base case for the recursion. The embedding involves three QRAM router qubits, one unused qubit, 
one routing qubit, and four data qubits. Note the distinction between a router qubit (in QRAM) and a routing qubit (for teleportation). This design ensures the root QRAM qubit can route to the border of the grid. Recursively, we can construct an embedding of $T_{m+2}$ into Grid$(2n+1,2n+1)$ as shown in Figure~\ref{fig:teleportation}. As for even-addressed width QRAMs, we can cut half the grid and make it a rectangular one with $M(2n+1,n)$ to embed $T_{m+1}$ QRAM into the 2D grid.

\subsection{Routing via Teleportation}
\label{subsec:teleport}

Our teleportation routing method is based on a technique called entanglement swapping \cite{pan1998experimental}, commonly used in the context of quantum repeater networks \cite{briegel1998quantum, van2014quantum}. If the intermediate qubits between two logical qubits are unused, they can be used as ancillae (routing qubits) to perform teleportation, as shown in Figure \ref{fig:teleportation}d and \ref{fig:teleportation}e. Local Einstein-Podolsky-Rosen (EPR) pair preparation and Bell State Measurement (BSM) are performed in parallel. As such, we can teleport a qubit over a long routing distance with a constant depth circuit. Each QRAM operation (e.g., remote \texttt{CSWAP}) can thus be implemented in $O(1)$ step, regardless of the routing distance.

As a result, our embedding from Sec~\ref{subsec:htree} is optimal in terms of routing latency. Since teleportation only introduces an $O(1)$ depth to each gate, the overall QRAM circuit depth remains in $O(\log M)$. Though H-tree is efficient enough to provide the optimal query latency, there are further optimizations are provided by \cite{opatrny2000embeddings,lin2003expansion,ducourthial1998implementing}, where improved versions of the H-tree recursion are found. As such, we can embed QRAM into a (constant-factor) denser grid.

\section{Noise-Robust Implementations}
\label{sec:noise}

QRAM is highly susceptible to noise and errors, which can cause information loss and reduce the fidelity of the stored states. Thus, QEM or QEC is essential for improving the reliability and accuracy of QRAM operations. This is critical for the success of many quantum algorithms and the development of fault-tolerant quantum computation. 

Prior work by \cite{hann} revealed intrinsic noise resilience in Bucket-Brigade QRAM by carefully analyzing the error propagation in QRAM circuits. We observed that virtual QRAM shares a similar biased-intrinsic noise resilience property, meaning that Z error in virtual QRAM is constrained to local qubits and will not propagate to the entire circuit even without any active quantum error correction or mitigation. To quantify the noise-resilience of virtual QRAM, we define \textit{query fidelity} for a single query $\ket{\psi_{in}}$ as 
$F = |\braket{\psi_{out}|\psi^{\prime}_{out}}|^2,$
where $\psi_{out}$ is the true output, and $\psi^{\prime}_{out}$ is the expected output. \par
With respect to this definition of query fidelity, we will show that in virtual QRAM, the query fidelity is lower bounded for an arbitrary $\ket{\psi_{in}}$. The infidelity is polynomially in terms of the address width m, rather than the overall tree size $2^m$. 

\subsection{Biased-Noise Resilience Analysis}
To quantify the noise-resilience of virtual QRAM, we define \textit{query fidelity} for a single query $\ket{\psi_{in}}$ as 
$F = |\braket{\psi_{out}|\psi^{\prime}_{out}}|^2,$
where $\psi_{out}$ is the true output, and $\psi^{\prime}_{out}$ is the expected output. \par
With respect to this definition of query fidelity, we will show that in virtual QRAM, the query fidelity is lower bounded for an arbitrary $\ket{\psi_{in}}$. The infidelity is polynomially in terms of the address width m, rather than the overall tree size $2^m$. 

\begin{figure}[t]
         \centering
         \includegraphics[width=0.9\linewidth]{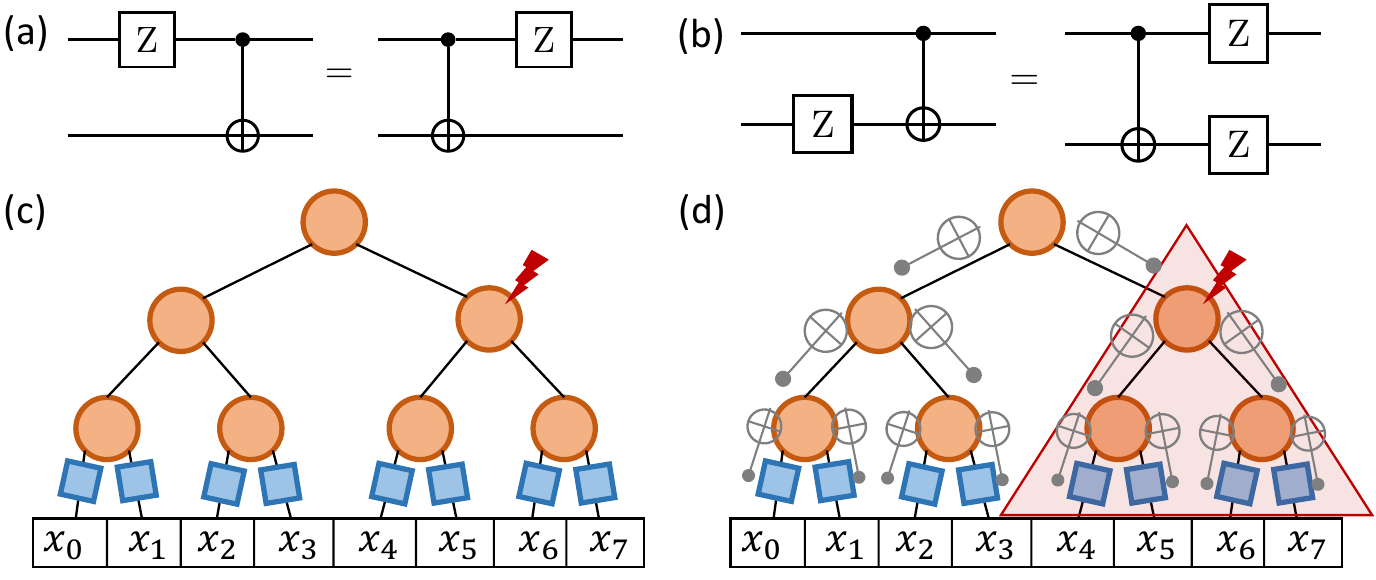}
         \caption{(a) and (b): Commutator relationship for Z and CNOT gates. (c) and (d): Z-error behavior in QRAM. A Z error only propagates to the subtree highlighted in red, due to the direction of the \texttt{CX} gates.}
         \label{fig:biasedz}\label{fig:zcom}
\end{figure}

To construct our Z-biased noise model, we assume that each qubit is subjected to the following phase-flip noise quantum channel, 
$\rho \rightarrow \rho' = (1-\epsilon) \rho + \epsilon Z\rho Z$.
Equivalently, a $Z$ error is applied to each qubit with probability $ \epsilon$.\ We show that in the presence of this qubit-based error channel, the QRAM part (beside SQC) in virtual QRAM has a lower bound in the query fidelity as
\begin{equation}
  F \ge 1 - 4 \epsilon  \log^2{M} =1 - 4 \epsilon  m^2
\end{equation}
where $m=\log(M)$ is the address width of QRAM.
We first present an outline of our methodology.\ Similar to the approach used in \cite{hann}, the locality behavior of the noise prevents the error from propagating throughout the entire QRAM, protecting the overall query fidelity. An example of this effect is illustrated in Fig.\ref{fig:zcom} — a Z error in the control qubit of a subsequent \texttt{CX} gate never propagates to the target qubit by commutator relationship of quantum gates. This property also holds in quantum routers using \texttt{CSWAP} gates, ensuring the error locality for the address loading stage. 

However, our virtual QRAM cannot prevent other Pauli errors, such as X and Y errors, from propagation. We will show that the fidelity under Z and X error channels has an exponential difference with respect to the QRAM width $m$.
\par
Consider the computational basis states $\ket{0}$, $\dots$, $\ket{2^{m+k} - 1}$ corresponding to different memory addresses. For a query $Q$, let $Q_{good} \subseteq \{0, \dots, 2^{m+k} - 1\}$ be the subset of $i$ such that $Q$ behaves ideally on $\ket{i}$: there are no $Z$ errors on any routers on the path of the branch corresponding to $i$. On the initial state $\ket{\psi_{in}} = \sum_{i = 0}^{2^{m+k} - 1} \alpha_i\ket{i}_A\ket{0}_B$,
 a query yields the state $\ket{\psi_{out}} = \left(\sum_{i \in Q_{good}}^{2^{m+k} - 1} \alpha_i\ket{i}_A\ket{x_i}_B\right) + \ket{\psi_{bad}}$  where $\ket{\psi_{bad}}$ is a super-position of all non-ideal branches (those with $Z$ errors).
 \begin{align*}
 F &=
 \biggr|\left\langle \psi_{out} \biggr|\sum_{i \in Q_{good}}^{2^{m+k} - 1} \alpha_i\ket{i}_A\ket{x_i}_B  \right\rangle + 
 \left\langle\psi_{out}\biggr |\psi_{bad} \right\rangle\biggr|^2\\
 &\ge \left(\sum_{i \in Q_{good}}|\alpha_i|^2 - \left(1 - \sum_{i \in Q_{good}}|\alpha_i|^2\right)\right)^2 \\
 &= \left(2\left(\sum_{i \in Q_{good}}|\alpha_i|^2 \right) - 1\right)^2
 \end{align*}
 where the inequality holds since $\ket{\psi_{out}}$ is normalized. Thus, it suffices to show that $Q_{good}$ is sufficiently large in expectation, i.e. most branches perform ideally. If $S$ and $S^\prime$ are sets of indices of the same size, they share equal possibility to coincide with $Q_{good}$, as our error model on the router is independent of the router itself. Letting $|Q_{good}| = c$, 
\begin{equation}
     \mathbb{E}\biggr[\sum_{i \in Q_{good}}|\alpha_i|^2\biggr] = \frac{\mathbb{E}[c]}{{2^m}} \implies \mathbb{E}[F] \ge \left(\frac{2\cdot\mathbb{E}[c]}{2^m} - 1\right)^2 
\end{equation}
where the expectation is taken over all possible errors. 
By the fact that the Z error will not propagate up across the tree to destroy other branches, the fulfillment condition for an ideal branch is that all the routers in the path are correct. Since each branch contains $m$ routers, the probability a branch behaves ideally is $(1-\epsilon)^{m^2}$, with $\mathbb{E}[c] = 2^m(1-\epsilon)^{m^2}$. 
Combining (3) and (4), we conclude that
$$1 - F \le 1 - \mathbb{E}[F]  \le 4\epsilon m^2 = 4\epsilon\log^2 M$$
where the final inequality is valid for $m \ge 1$, $\epsilon \ge 0$. This proves the fidelity bound in Equation (3).

Precisely, if dual-rail encoding introduced in Section \ref{sec:arch} is adopted, the router qubits and data qubits are duplicated, which doubles the errors of each router in the circuit. The above derivation, however, is still valid because the locality of the Z error behavior is not relevant to the choice of encoding. Using the same methodology, we arrive at a bound with only a constant factor difference for dual-rail-encoding virtual QRAM:
$F_{\mathrm{dual-rail}}\geq 1 - 8 \epsilon m^2.$

On the contrary, the circuit has no noise-resilience property for X errors. Any single X error will propagate to the root qubit of the QRAM, leading to a complete destruction of the query state. Thus to achieve an ideal state, it suffices to show that all the qubits in QRAM are correct. As such, for X error channel with error prob $\epsilon$, the lower bound of infidelity is $1 - 8 \epsilon m\cdot 2^m$, exponential in the total number of qubits. Similar to the X error behavior in the QRAM part, any single Pauli error in SQC is fatal for the query fidelity. Consequently, for an SQC width k, the query fidelity under arbitrary Pauli errors with error rate $\epsilon$ is lower bounded by $1 -  (\epsilon k\cdot 2^k)$.
Combining these bounds, We conclude the virtual QRAM with QRAM address width m and SQC width k will lower bound the overall query fidelity as
\begin{align}
    F_{\textrm{virtual},Z}\geq 1 - 8 \epsilon (m+1)  \cdot 2^k(k+m)
    \label{x}\\
    F_{\textrm{virtual},X}\geq 1 - 8 \epsilon (m+1)\cdot 2^k(k+2^m)
    \label{z}
\end{align}
for Z errors and X errors, respectively.
Additionally, our biased-noise analysis can be easily extended to a gate-based error channel with errors randomly applied using Monte Carlo sampling to quantum gates, up to a constant factor difference. By the observation that each branch of the QRAM intersects with at most $O(m)$ gates, the lower bound of the fidelity has the same asymptotic scaling as under the qubit-based noise model.

\subsection{Asymmetric Error Correction}
In this section, we explore the fault-tolerant implementation of the virtual QRAM in future quantum hardware, using rectangular surface code to combine error correction design with intrinsic QRAM noise resilience. We assume that the error rate of physical qubits is unbiased with respect to both X and Z, whereas a logical qubit generated by the rectangular surface code exhibits a biased error rate. Adopting this biased-error surface code as qubits in virtual QRAM, balanced fidelity for X and Z errors can be achieved.

Based on the different lower bounds of query fidelity under X and Z error channels, we need a careful choice of the surface code for the different parts in virtual QRAM. First, the logical error rate ratio of X and Z is related to the physical error rate, surface code threshold, and the code distances $dx$ and $dz$ \cite{bonilla2021xzzx}:
$\frac{p_xl}{p_zl}\approx\left(\frac{p}{p_{th}}\right)^{dx-dz}$.
To balance the logical error rates of X and Z error channels, we adopt the bound from equation \ref{x} and \ref{z} and let $F_x=F_z$, then $\frac{\epsilon_x}{\epsilon_z} = \frac{p_xl}{p_zl} =\frac{(m+k)\cdot2^k}{2^k(k+2^m)}=\frac{k+m}{k+2^m}$.
We obtain the strategy of designing rectangular surface code for each physical qubit in QRAM is choosing lengths of the surface code  $dx$ and $dz$ as:
\begin{align}
    dx-dz \approx \frac{\log(\frac{k+m}{k+2^m})}{\log(\frac{p}{p_{th}})}
\end{align}
Since the SQC does not have biased-noise resilience, we can encode $k$ address qubits using regular square surface code to achieve full protection for the entire virtual QRAM.

\section{Evaluation Methodology}
\label{sec:eval}

\subsection{Baseline Architectures}
We theoretically analyze the performance of the new circuit and perform simulation in comparison to two baseline architectures, BB (Bucket-Brigade QRAM) and SS (Select-Swap QRAM), described in Sec \ref{sec:background}. These two QRAM architectures present state-of-the-art QRAM architecture designs in both consumption of quantum resources and performance of quantum circuits including circuit depth and noise-resilience properties. 

\subsection{Feynman Path Simulation}
We utilized a classical simulation technique called Feynman-Path Simulation (FPS) to efficiently compute the result of QRAM queries. In FPS, each memory address corresponds to one path. Despite path number and running time scaling exponentially in the address space, FPS can still efficiently simulate and analyze larger QRAMs, on account of the following desirable property. QRAM circuits are constructed from a small, fixed set of classical-reversible gates, meaning that all these gates do not map a single computational basis state to a superposition over basis states. As a consequence, the storage overhead 
does not increase exponentially in the depth of the circuit,
and instead remains constant. Notably, Pauli gates have the same property to ensure we execute noisy QRAM simulations with negligible overhead.
Compared to the special-purpose simulator in \cite{hann}, our Feynman Path implementation is the first general-purpose QRAM simulation that is capable of handling arbitrary input (e.g., memory capacity, address state, and noise models). 
\par

\subsection{Experimental Setup}

Noise-free or Pauli-noise circuit simulations are implemented with a single core on a single node, with the largest simulations using 1.5 MB of RAM. For each value of QRAM address width $m$, we execute $1024$ shots to achieve the average fidelity, using a gate-based error model applied via Monte Carlo sampling. We assume devices with 2D square grid connectivity, commonly used in NISQ or FTQC architectures.

\section{Results}
\label{sec:results}

\subsection{QRAM Resource Estimation}
\begin{table*}[t]
    \centering
    \begin{tabular}{ p{3.8cm}||p{2.3cm}|p{2.3cm}|p{2.3cm}|p{2.3cm}|p{2.3cm} }
     \hline
     \hline
       & RAW & OPT: 1 & OPT: 2 & OPT: 3 & OPT: ALL \\
     \hline
     Qubits (bit encoding)   & $6\cdot2^m+k$   & $\mathbf{4\cdot2^m+k}$  & $6\cdot2^m+k$ & $6\cdot2^m+k$ & $\mathbf{4\cdot2^m+k}$ \\
     \hline
     Circuit depth & $m^2+(m+1)\cdot2^k$ & $m^2+(m+1)\cdot2^k$  & $m^2+(m+1)\cdot2^k$ &$\mathbf{m+(m+1)\cdot2^k}$ &$\mathbf{m+(m+1)\cdot2^k}$\\
     \hline
     Classical controlled gates & $2^{m+k-1}$ &  $2^{m+k-1}$  & $\mathbf{2^{m+k-2}}$ & $2^{m+k-1}$& $\mathbf{2^{m+k-2}}$ \\
     \hline
     \hline
    \end{tabular}
    \caption{Resource overhead improvements from three key optimization methods in Section \ref{sec:arch}.}
    \label{tab:optimization}
\end{table*}
\begin{table}[t]
    \centering
    \begin{tabular}{ p{1.95cm}||p{1.62cm}|p{1.64cm}|p{1.64cm} }
     \hline
     \hline
      & SQC+BB & SQC+SS & Our QRAM\\
      \hline
     Qubits   & $2^m+k$   & $2^m+k$ & $2^m+k$\\
     \hline
     Circuit depth & $m\cdot2^k$ &  $m^2\cdot2^k$  & $m\cdot2^k$ \\
     \hline
     T count & $(2^m+k)\cdot2^k$ &  $2^m+k\cdot2^k$  & $2^m+k\cdot2^k$  \\
     \hline
     T depth & $(m+k)\cdot2^k$ &  $m+k\cdot2^k$  & $m+k\cdot2^k$ \\
     \hline
     Clifford depth & $(m+k)\cdot2^k$ &  $(m^2+k)\cdot2^k$  & $(m+k)\cdot2^k$ \\
     \hline
     \hline
    \end{tabular}
    \caption{Resource overhead comparison between different implementations of virtual QRAM. All costs are in Big-O.}
    \label{tab:individualcounting}
\end{table}
The improvements breakdown by different optimization techniques in Section \ref{sec:arch} are listed in Table \ref{tab:optimization}. Table \ref{tab:individualcounting} presents a comprehensive comparison of multiple current quantum architectures, including concrete parameters such as circuit depth, number of qubits, and T gate depth, among others. Notably, the asymptotic scaling in the table indicates that all three architectures have the same scaling with respect to qubit count. SQC+BB (Baseline B) is a load-multiple-times architecture that suffers from deficiencies in exponential $O(2^K)$ overhead in T depth and T counting complexity. SQC+SS (Baseline S) is a load-once architecture; however, its swap-network is not as efficient as the router-based QRAM, since it lacks a pipelining strategy to load the address discussed in Section \ref{sec:arch}. Consequently, the circuit depth and Clifford depth of SQC+SS will be quadratically larger than our new QRAM architecture with factor $O(m^2)$. Therefore, our virtual QRAM \textit{outperforms} or at least \textit{matches} any resource counting compared with the state-of-the-art QRAM architectures.


\subsection{Mapping and Routing Overhead}
\begin{figure}[t]
         \centering
         \includegraphics[width=0.8\linewidth]{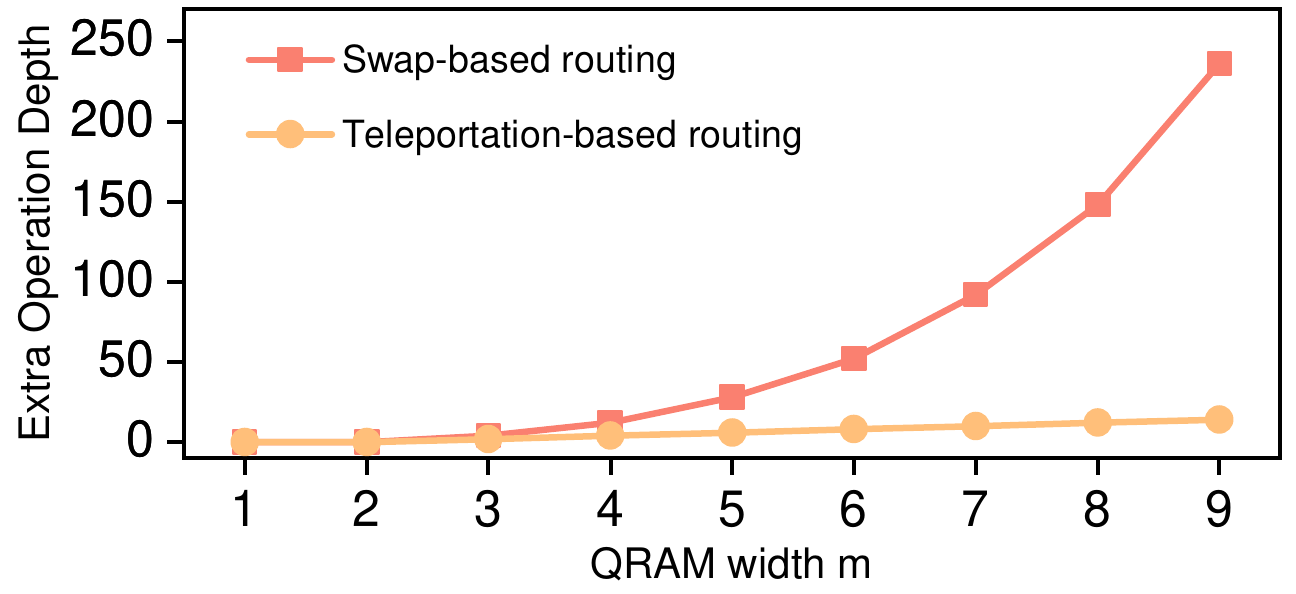}
         \caption{Additional operations after mapping in 2D nearest-neighbor architectures. Swap-based communication scales exponentially worse than teleportation-based communication.}
         \label{fig:swapdepth}
\end{figure}

Figure \ref{fig:swapdepth} presents the results of our constructive mapping strategy, counting the extra operation depth induced from connectivity constraints in NISQ hardware. The conventional swap-based routing,  when applied to QRAM, leads to exponential extra \texttt{SWAP} depth overhead and loss of the logarithmic scaling of query depth in terms of QRAM width m. In contrast, the teleportation-based routing consistently outperforms the other, exhibiting an exponential advantage in extra operation depth. 
Figure \ref{fig:swapdepth} also indicates that QRAM under embedding only introduces a linear overhead for circuit depth, which protects the query latency to be \textit{unchanged} under qubit mapping and routing. 
Meanwhile, our H-tree embedding and the teleportation scheme reduce the circuit depth exponentially but with only constant qubit resource overhead. With $m=2n$ address qubits, unused qubits occupy $25\%$ of the total qubits: $(2^n-1)^2$ vs $(2^{(n+1)}-1)^2$. The proportion of unused qubits can be further reduced by improved tree embedding discussed in Sec~\ref{subsec:teleport}.

\begin{figure}[t]
         \centering
         \includegraphics[width=\linewidth]{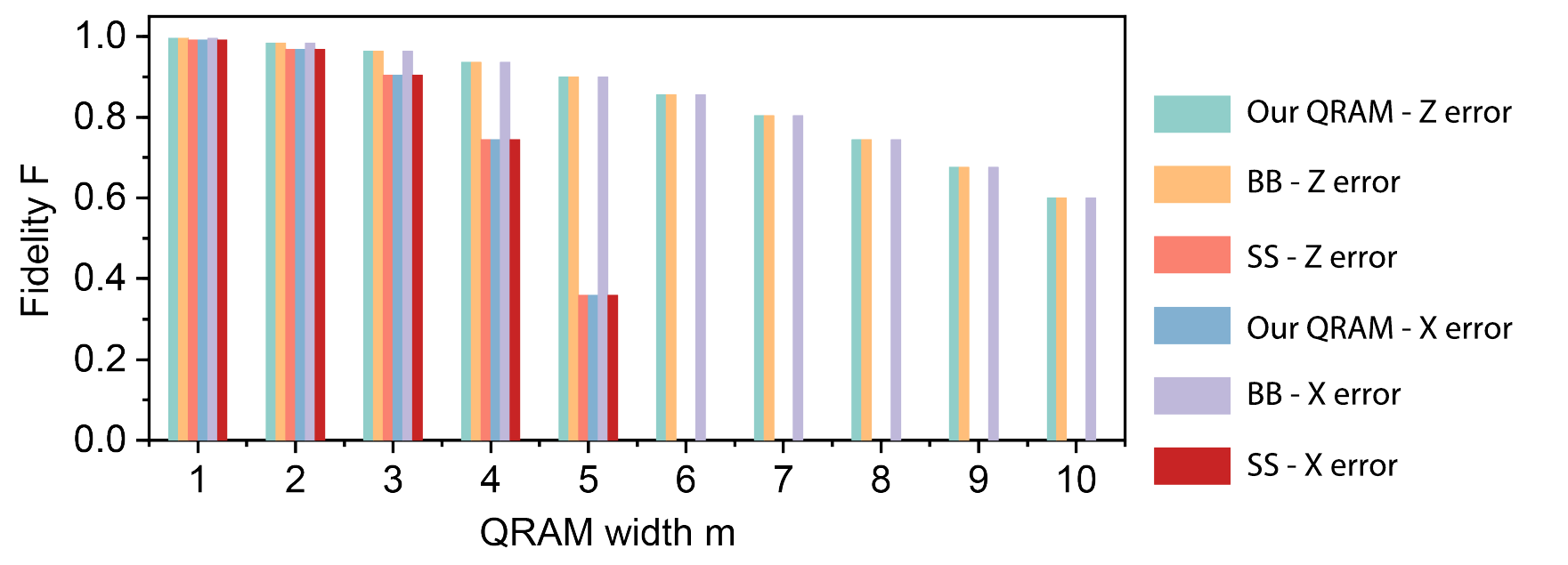}
         \caption{Fidelity comparison for different QRAM architectures. We observe fidelity decays polynomially for Z errors in virtual QRAM and BB QRAM, but for X errors only in BB QRAM.}
         \label{fig:comparez}
\end{figure}

\begin{figure}[t]
         \centering
     \begin{subfigure}[hb]{0.47\linewidth}
         \centering
         \includegraphics[width=\linewidth]{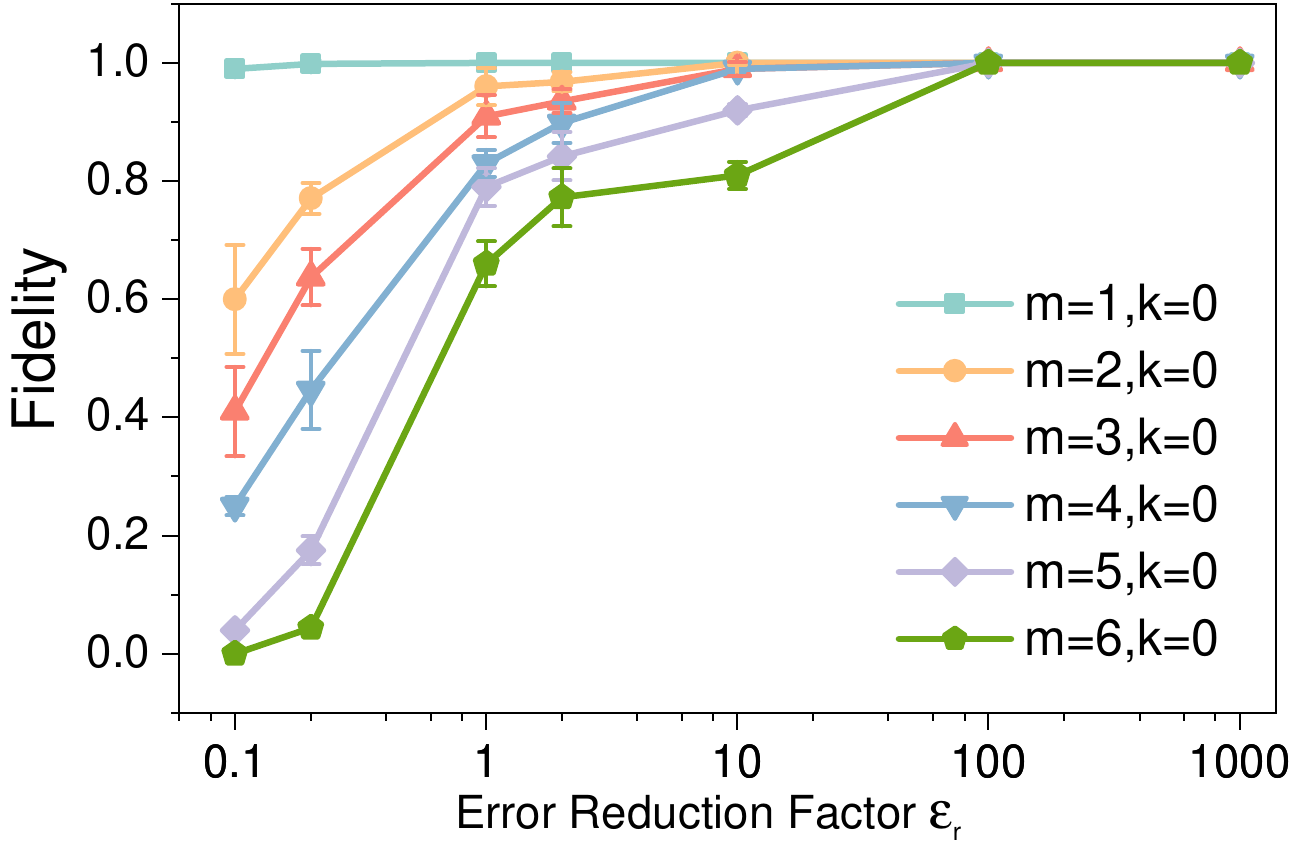}
     \end{subfigure}
     \hfill
     \begin{subfigure}[hb]{0.47\linewidth}
         \centering
         \includegraphics[width=\linewidth]{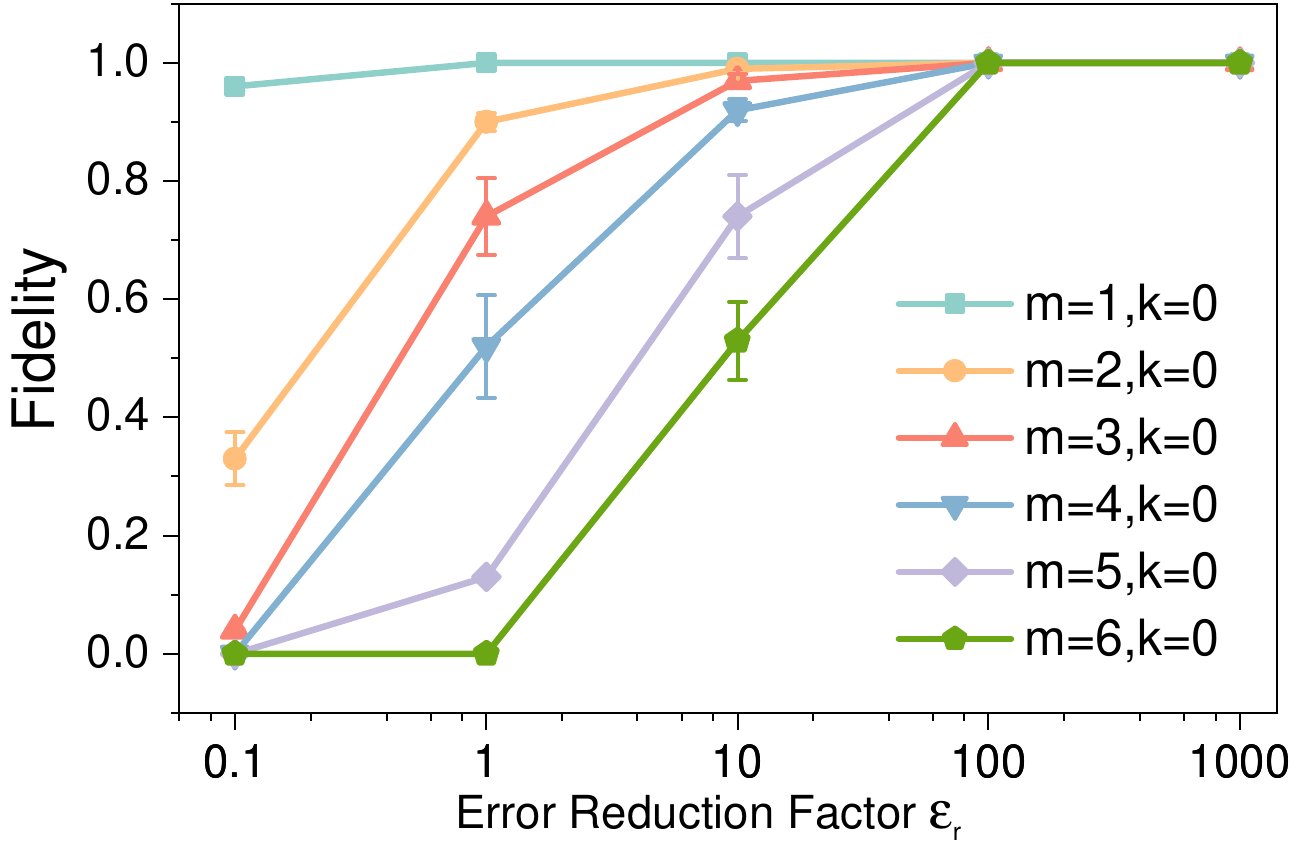}
     \end{subfigure}
     \caption{(Left) Fidelity with phase flip error model by numerical simulation. (Right) Fidelity with bit flip error model by numerical simulation.} 
      \label{fig:simulation}
\end{figure}

\begin{figure}[t]
      \centering
     \includegraphics[width=0.85\linewidth]{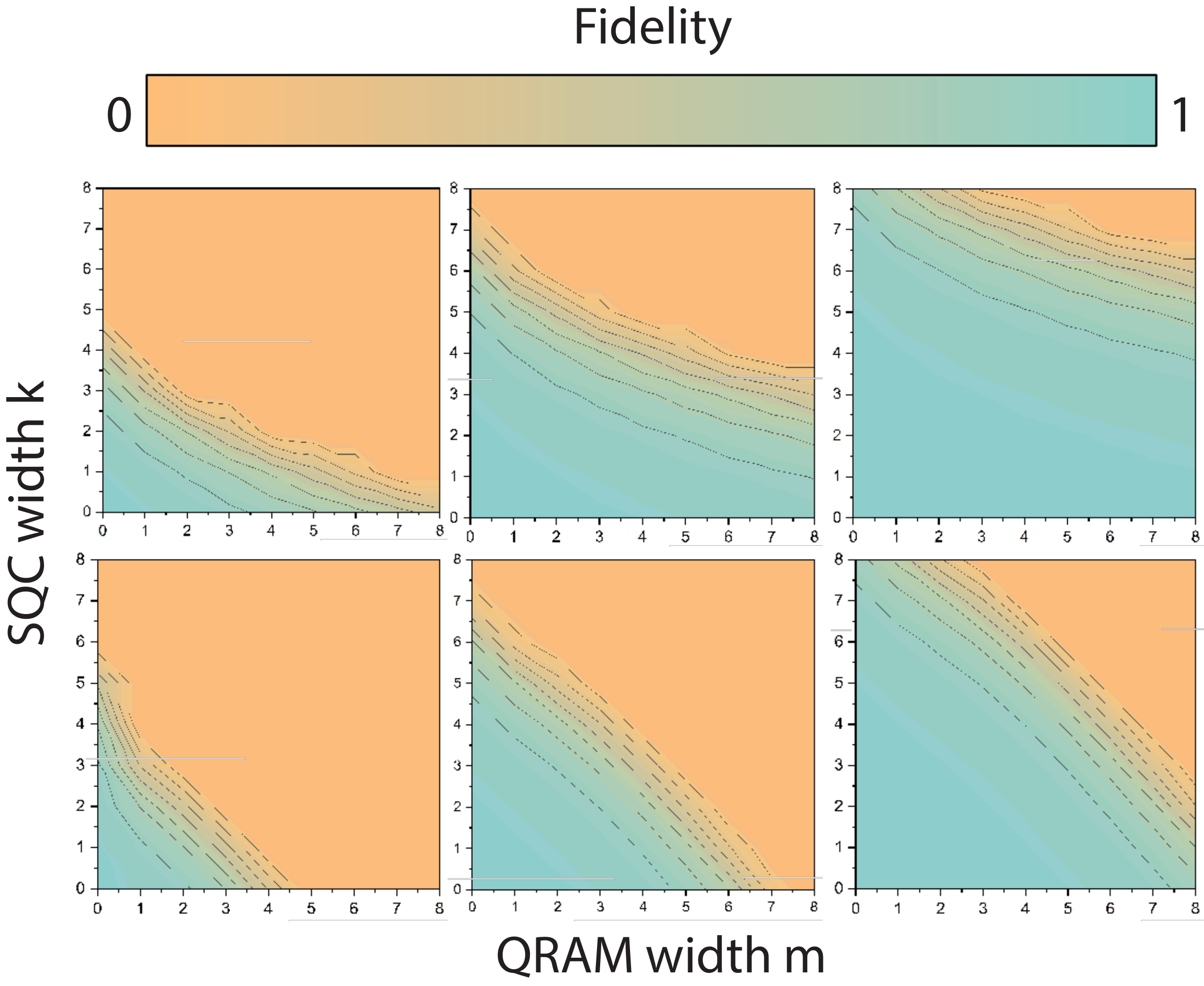}
     \caption{Virtual QRAM fidelity scaling under X and Z error. Top/Bottom: Z/X error with (Left) $\epsilon_r=1$, (Middle) $\epsilon_r=10$, (Right) $\epsilon_r=100$.}
    \label{fig:scaling}
\end{figure}

\subsection{Biased-noise simulation}
As illustrated in Section \ref{sec:noise}, the QRAM architecture is robust to Z-biased error. We further validate this property by calculating the virtual QRAM fidelity and comparing it to Baseline B, which is robust to arbitrary error channels but has the same scaling as the new virtual QRAM for Z error. As expected, Baseline S exhibits no noise resilience. Figure \ref{fig:comparez} shows the scaling of the fidelity under Pauli X and Z errors for different architectures with error rate $\epsilon=10^{-3}$.

Moreover, our simulations in Figure \ref{fig:simulation} demonstrate the fidelity gap between Z-biased noise and X-biased noise, with far better performance in the former. We also provide Figure \ref{fig:scaling} to illustrate the trade-off between the QRAM width m and SQC width k under the single qubit Z and X error model. Our plot indicates that the fidelity decays \textit{exponentially} faster when increasing the SQC width parameter $k$ than the QRAM width $m$.

\section{Related Works}\label{sec:related}
Recently a similar architecture design for QRAM by Chen et al. designed a ``load-once'' QRAM architecture capable of querying memory with more than 1 bit in each memory cell, i.e., a data width of $k\geq 1$ \cite{chen2023efficient}. The difference between theirs and our work is that we target creating larger address widths rather than simply repeating the data retrieved multiple times to query the same address for more bus qubits. Notably, some previous works, e.g., Connor et al. \cite{hann}, mentioned similar strategies to generalize to high data width. Our virtual QRAM is compatible with a data width larger than $1$ by repeatedly querying memory cells one bit at a time, by taking advantage of the parallel retrieval from \cite{chen2023efficient} in our virtual QRAM.

During the preparation of this work, Jaques and Rattew made similar claims on qubit mapping in QRAM in a recent paper \cite{jaques2023qram}. One major limitation of QRAM highlighted in both our work and \cite{jaques2023qram} is the signal latency for communicating qubits within the QRAM. \cite{jaques2023qram} assumes a quantum bus line for communication, which has latency linear to its distance, while our work proposes a novel teleportation-based routing scheme to overcome this bottleneck. Ultimately, via teleportation, we can propagate information across QRAM qubits faster (only limited by classical communication or the speed of light), with negligible quantum delay. This is an important step towards achieving quantum advantage for QRAM with relatively small query delay.

There are significant advances in hardware development towards QRAM using superconducting devices and cold atom arrays. For example, the key element of deterministic \texttt{CSWAP} operations between superconducting cavities have been demonstrated in previous work \cite{gao2019entanglement, xue2023hybrid}. \texttt{CSWAP} operations using superconducting transmon devices have also been demonstrated \cite{miao2023implementation}. \texttt{Toffoli} gate and \texttt{CSWAP} for QRAM have also been investigated using the Rydberg blockade.

\section{Conclusion}

The use of QRAM is ubiquitous in quantum algorithms. A successful implementation of practical QRAMs could unlock the full potential of quantum computing and bring us closer to realizing practical applications such as optimization, machine learning, and cryptography. Our proposed QRAM architecture addresses the challenges of memory capacity, query latency, and fault-tolerance through innovations in virtualizing QRAM, latency-free mapping to 2D grid architecture, and leveraging intrinsic biased-noise resilience in the circuits. We have shown an end-to-end systems architecture for performing high-fidelity queries of large memories, and identified key technology advances (such as gate error rate reduction or error correction code distance) needed to scale up QRAM and quantum computing platforms as a whole.


\begin{acks}
YD acknowledges support from NSF (under award CCF-2312754) and Yale University. SMG was supported by the Air Force Office of Scientific Research under award number FA9550-21-1-0209. We thank Liang Jiang, Shruti Puri, Junyu Liu, Nathan Wiebe, and Daniel Weiss for fruitful discussions.
\end{acks}

\bibliographystyle{ACM-Reference-Format}
\bibliography{main}

\appendix
\label{sec:appendix}
\section{Is QRAM Viable on Current QPUs?}

\begin{figure}[t]
    \centering
    \includegraphics[width=0.8\linewidth]{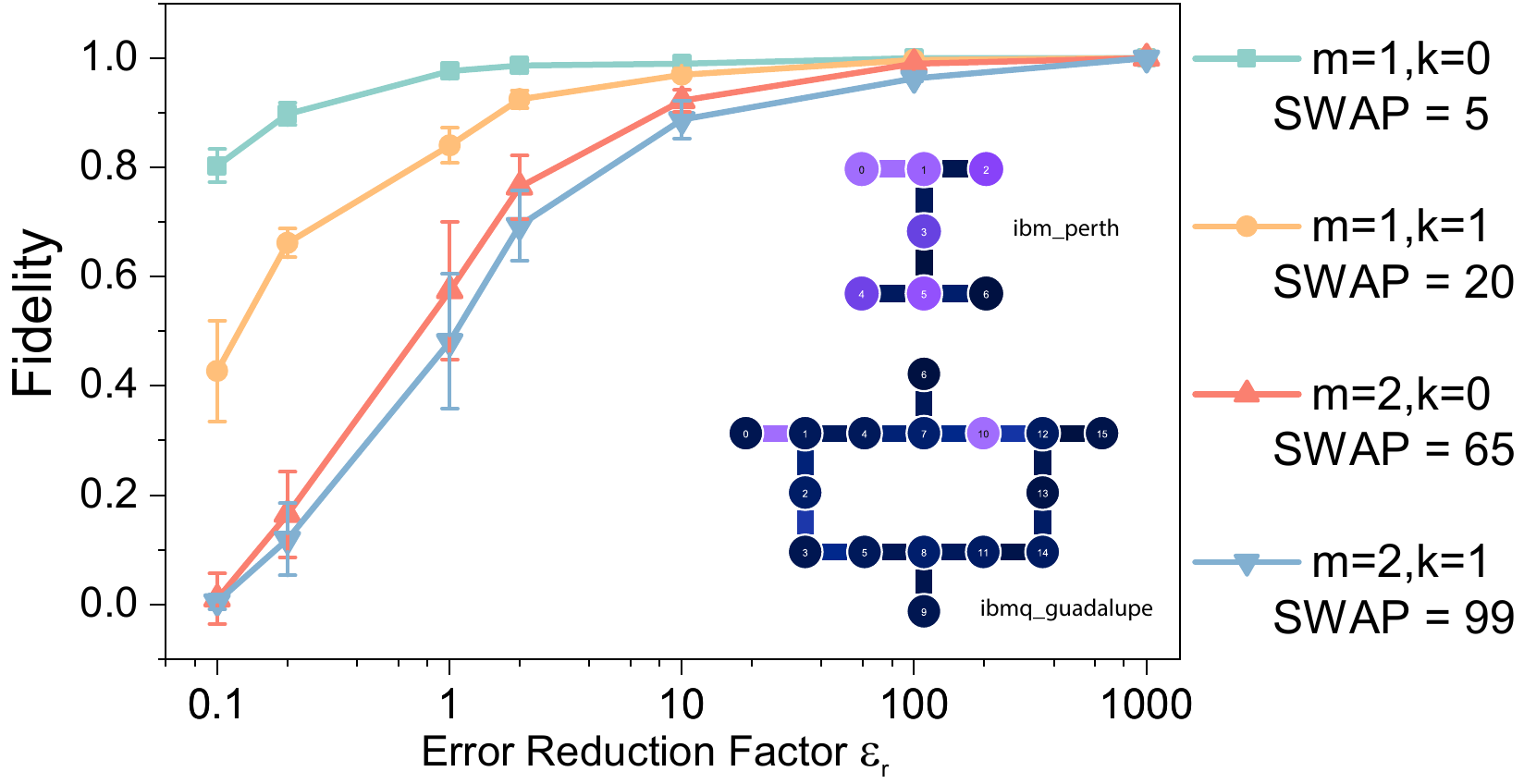}
    \caption{Fidelity with reduced error model from IBMQ. Extra \texttt{SWAP} gates are reported under the legend.}
    \label{fig:ibmq}
\end{figure}

Current QPUs have limited qubit count, device connectivity, and fidelity. In this section, we investigate the main technological improvements needed to implement QRAM. Our real hardware noise simulation experiments are implemented in Python 3.9, interfacing with IBM's Qiskit software library. All the IBMQ noisy simulation experiments are executed with 200 shots, using 48 cores and 30GB of RAM with error models from IBM's $ibm\_perth$ for $m=1$ cases and from IBM's $ibmq\_guadalupe$ for $m=2$ cases. The topology of the two machines is listed in Figure \ref{fig:ibmq}. 
Our analysis in Section \ref{sec:noise} reveals a lower bound of the fidelity that suggests high fidelity can be achieved for small-scale QRAM by enhancing the error rate of the quantum hardware. To this end, we introduce an \textit{Error Reduction Factor} ($\epsilon_r$) that describes the error rate ratio between expected future hardware and current hardware. $\epsilon_r=\frac{\text{current error rate}}{\text{future error rate}}$, with the current hardware error rate assumed to be $10^{-3}$.

In Figure \ref{fig:ibmq}, we assess the fidelity of the small-scale virtual QRAM under a realistic noise model obtained from IBM quantum hardware and simulate the virtual QRAM with error rates $\epsilon_r$ times lower than the current model to predict its performance in the future machine. Since IBM hardware has sparser connectivity than 2D square grid analyzed in Sec~\ref{sec:mapping}, extra \texttt{SWAP} gates by Qiskit default transpiler Sabre \cite{li2019tackling} are needed to implement QRAM operations, with counting in Figure \ref{fig:simulation}. Despite the extra \texttt{SWAP} operations, encouragingly, our results indicate that with a future noise model \textit{10 times} better than the current technology, we can achieve significantly improved query fidelity. Furthermore, when error rates are reduced to $10^{-5}$ possibly via near-term small-scale error correction, then the query fidelity can reach higher than 0.98.

\end{document}